\documentclass[a4paper,11pt]{article}

\usepackage[T1]{fontenc}
\usepackage[utf8]{inputenc}
\usepackage[american]{babel}
\usepackage{microtype}
\usepackage{amsmath}
\usepackage{amssymb}
\usepackage{amsthm}
\usepackage{mathpazo}
\usepackage[numbers,sort&compress]{natbib}
\usepackage{doi}
\usepackage{booktabs}
\usepackage[margin=2.5cm]{geometry}
\usepackage{hyperref}
\usepackage{spotltl}
\usepackage{xspace}

\newtheorem{example}{Example}
\newcommand{\alv}{\mathit{alive}}

\newcommand{\phiB}{\varphi_{\!B}}
\newcommand{\phiG}{\varphi_{\!G}}
\newcommand{\phiS}{\varphi_{\!S}}
\newcommand{\phiO}{\varphi_{\!O}}

\newcommand{\LTLB}{\mathit{LTL}_{\!B}}
\newcommand{\LTLG}{\mathit{LTL}_{\!G}}
\newcommand{\LTLS}{\mathit{LTL}_{\!S}}
\newcommand{\LTLO}{\mathit{LTL}_{\!O}}
\newcommand{\LTLf}{LTL$_{\!f}$\xspace}
\newcommand{\LTLfB}{\mathit{LTL}_{\!\textsf{f}B}}
\newcommand{\btriangle}{\mathbin{\triangle}}

\newcommand{\sat}{\vDash}
\newcommand{\satf}{\vDash_{\!\mathrm{fin}}}

\newcommand{\eqalv}{\equiv_{\!\mathrm{fin}}}

\newcommand{\tv}{\mathsf{t}}
\newcommand{\tB}{\mathsf{t}_{\mathsf{B}}}
\newcommand{\tG}{\mathsf{t}_{\mathsf{G}}}
\newcommand{\tS}{\mathsf{t}_{\mathsf{S}}}
\newcommand{\tO}{\mathsf{t}_{\mathsf{O}}}

\title{A note on the reduction from \texorpdfstring{\LTLf}{LTLf} to LTL}
\author{Alexandre Duret-Lutz}
\date{2026-08-31}
\begin{document}
\maketitle

\begin{abstract}
  \LTLf, a finite word variant of LTL, can be reduced to LTL by
  introducing a new atomic proposition indicating the prefix of the
  infinite words that correspond to the finite words that the original
  \LTLf formula was considering.  Such a reduction was originally
  proposed by \citet{degiacomo.13.ijcai}.

  However, while any LTL formula reduced from \LTLf describes an
  \emph{obligation property} in the hierarchy of \citet{manna.87.podc},
  the aforementioned reduction does not provide an LTL formula that
  belongs to the \emph{syntactic obligation} fragment of LTL.

  This note shows how the reduction was fixed in Spot in order to
  ensure that the resulting LTL formula is always a syntactic
  obligation.  Doing so allows algorithms specialized to syntactic
  obligation to be used on \LTLf formulas.  For instance, in previous
  work~\cite{duret.26.cav} we described a specialized translation from
  syntactic obligations to minimal, weak, deterministic Büchi automata
  that would not be usable with the original reduction.
\end{abstract}

\section{Preliminaries}
\label{sec:prelim}

We assume that the reader is familiar with Linear Temporal Logic (LTL)
over infinite words and with \LTLf, one of its restrictions to finite
words~\cite{degiacomo.13.ijcai}.  We use the standard temporal
operators $\F$ (eventually), $\G$ (always), $\X$ (weak next),
$\StrongX$ (strong next), $\U$ (until), $\W$ (weak until), $\R$
(release), and $\M$ (strong release).  In \LTLf all operators are
interpreted on nonempty \emph{finite} words: strong operators
($\F$, $\U$, $\M$) have to fulfill their promise before the end of the
finite word, $\X f$ is vacuously true at the last position, and
$\StrongX f$ requires a next position.

\paragraph{Satisfaction.}
We write $(w, i) \sat \varphi$ to mean that the infinite word $w$ satisfies
the LTL formula $\varphi$ at position $i \geq 0$, and $w \satf f$ to mean
that the finite word $w$ satisfies the \LTLf formula $f$ (at its first
position).

\paragraph{Syntactic hierarchy.}
We recall the syntactic hierarchy of~\citet{manna.87.podc}.
The hierarchy classifies LTL formulas into
classes (Safety: $\varphi_S$; Guarantee: $\varphi_G$; Obligation: $\varphi_O$; Persistence; Recurrence; Reactivity) that correspond to automata-theoretic classes.
The precise definition of the syntactic sub-classes most relevant to this
paper is given by the following grammars, where $v$ ranges over atomic
propositions:
\begin{align*}
  \phiB ::={}&
      \bot \mid \top \mid v
      \mid \lnot\phiB
      \mid \phiB \land \phiB
      \mid \phiB \lor  \phiB
      \mid \phiB \to   \phiB
      \mid \phiB \liff \phiB
      \mid \phiB \lxor \phiB
      \mid \X\phiB \\
  \phiG ::={}&
      \phiB
      \mid \lnot\phiS
      \mid \phiG \land \phiG
      \mid \phiG \lor  \phiG
      \mid \phiS \to   \phiG
      \mid \X\phiG
      \mid \F\phiG
      \mid \phiG \U \phiG
      \mid \phiG \M \phiG \\
  \phiS ::={}&
      \phiB
      \mid \lnot\phiG
      \mid \phiS \land \phiS
      \mid \phiS \lor  \phiS
      \mid \phiG \to   \phiS
      \mid \X\phiS
      \mid \G\phiS
      \mid \phiS \R \phiS
      \mid \phiS \W \phiS \\
  \phiO ::={}&
      \phiG \mid \phiS
      \mid \lnot\phiO
      \mid \phiO \land \phiO
      \mid \phiO \lor  \phiO
      \mid \phiO \to   \phiO
      \mid \phiO \liff \phiO
      \mid \phiO \lxor \phiO
      \mid \X\phiO \\
      &\mid \phiO \U \phiG
       \mid \phiO \R \phiS
       \mid \phiS \W \phiO
       \mid \phiG \M \phiO
\end{align*}
Let $\LTLB$, $\LTLG$, $\LTLS$, and $\LTLO$ denote the sets of LTL formulas
generated by $\phiB$, $\phiG$, $\phiS$, and $\phiO$, respectively.
Intuitively:
\begin{itemize}
  \item $\LTLB$ (Bottom) contains propositional formulas
        plus $\X$.
  \item $\LTLG$ (Guarantee) contains no weak operators
        ($\G$, $\W$, $\R$).
  \item $\LTLS$ (Safety) contains no strong operators
        ($\F$, $\U$, $\M$).
  \item $\LTLO$ (Obligation) is the closure of $\LTLG \cup
        \LTLS$ under Boolean connectives, $\X$, and four \emph{mixed}
        temporal operators: $\phiO \U \phiG$, $\phiO \R \phiS$,
        $\phiS \W \phiO$, and $\phiG \M \phiO$.  These additional grammar
        rules are due to \citet{chang.92.icalp}.
\end{itemize}
Note that $\LTLG$ and $\LTLS$ are \emph{not} closed under $\lxor$ and
$\liff$; however, $\LTLB$ and $\LTLO$ are.

We will also need a variant of $\LTLB$ that includes the strong-next
operator $\StrongX$ of \LTLf.  Let $\LTLfB$ denote the set of \LTLf
formulas generated by $\phiB$ extended with the production
$\mid \StrongX\phiB$.

\paragraph{The alive encoding.}
Following~\citet{degiacomo.13.ijcai}, a nonempty finite word
$w = w_0 \cdots w_{n-1}$ over an alphabet $2^{AP}$ can be encoded as
an infinite word $w' = w_0' w_1' \cdots$ over $2^{AP \cup \{\alv\}}$
where
\[
  w_i' = w_i \cup \{\alv\}\ \text{ for } i < n,
  \qquad
  \alv \notin w_i'\ \text{ for } i \geq n.
\]
The proposition $\alv$ marks the \emph{live} prefix of the infinite
word; positions $i \geq n$ are \emph{dead}.


\paragraph{Equivalence at alive positions.}
We say that two LTL formulas $\psi_1$ and $\psi_2$ are
\emph{equivalent at alive positions}, written $\psi_1 \eqalv \psi_2$, if
for every infinite word $w'$ conforming to the alive encoding and every
alive position $i$ of $w'$, $(w', i) \sat \psi_1$ if and only if
$(w', i) \sat \psi_2$.

\section{The Original Reduction from \texorpdfstring{\LTLf}{LTLf} to LTL}
\label{sec:original}

\citet{degiacomo.13.ijcai} define $\tv$ and prove that, for any \LTLf
formula $f$ and any nonempty finite word $w$, if $w'$ is the alive
encoding of $w$, then
\begin{equation}\label{eq:wrapper-dv}
  \mathrm{from\_ltlf}(f) \;=\; \tv(f) \;\land\; \alv \;\land\; (\alv \U \G\lnot\alv)
\end{equation}
is satisfied by $w'$ if and only if $w \satf f$.  Consequently,
$\mathrm{from\_ltlf}(f)$ is satisfiable by an alive-encoded word if and only if
$f$ is satisfied by some nonempty finite word.
The function $\tv$ is defined by the following equations.  For Boolean
operators and atomic propositions, $\tv$ is the identity (applied
homomorphically).  For temporal operators, live-prefix constraints are
injected:
\begin{alignat*}{2}
  \tv(\bot)       &= \bot,\quad \tv(\top) = \top,\quad \tv(v) = v \\
  \tv(\lnot f)    &= \lnot\tv(f) \\
  \tv(f \odot g) &= \tv(f) \odot \tv(g)\quad\text{for any}~\odot \in \{\land,\lor,\to,\liff,\lxor\}
                   \\
  \tv(\StrongX f) &= \X(\alv \land \tv(f))
                   &&\quad\text{[strong next: target must be alive]} \\
  \tv(\X f)       &= \X(\lnot\alv \lor \tv(f))
                   &&\quad\text{[weak next: trivially true past the end]} \\
  \tv(\F f)       &= \F(\alv \land \tv(f))
                   &&\quad\text{[eventuality must land alive]} \\
  \tv(\G f)       &= \G(\lnot\alv \lor \tv(f))
                   &&\quad\text{[universality trivially holds at dead positions]}\\
  \tv(f \U g)     &= \tv(f) \U (\alv \land \tv(g))   \\
  \tv(f \R g)     &= \tv(f) \R (\lnot\alv \lor \tv(g)) \\
  \tv(f \W g)     &= (\lnot\alv \lor \tv(f)) \W \tv(g) \\
  \tv(f \M g)     &= (\alv \land \tv(f)) \M \tv(g)
\end{alignat*}
Strong operators ($\F$, $\U$, $\M$, $\StrongX$) require the target
position to be alive; weak operators ($\G$, $\R$, $\W$, $\X$) allow a
dead position as a trivial escape.

The definition above was used to implement $\mathrm{from\_ltlf}$ in
Spot~\cite{duret.22.cav} from version 2.2.2 to version 2.15.1.

\paragraph{The translation is not a syntactic obligation.}
Using $\mathrm{from\_ltlf}$, every \LTLf formula translates to an LTL
formula whose language is an obligation
property in~\cite{manna.87.podc} hierarchy.  However, the output of
$\mathrm{from\_ltlf}$ is generally \emph{not} in $\LTLO$.  There are
two issues.

The first issue is the wrapper itself: syntactically,
$\alv \U \G\lnot\alv$ is a persistence formula, not in $\LTLO$.  This
is a minor problem and is easily fixed: on alive-encoded words,
$\alv \U \G\lnot\alv$ is equivalent to
$\F\lnot\alv \land (\alv \W \G\lnot\alv)$, which is in $\LTLO$.

The second, deeper issue is that even the inner formula $\tv(f)$ is
generally not in $\LTLO$.

\begin{example}\label{ex:gfa}
  Consider $\tv(\G(\F a)) = \G(\lnot\alv \lor \F(\alv \land a))$,
  or its dual $\tv(\F(\G a)) = \F(\alv \land \G(\lnot\alv \lor a))$.
  Neither result is in $\LTLO$.
\end{example}

The failure in Example~\ref{ex:gfa} is structural: $\tv$ uses the same
recursive call in both the ``$\phiG$-context'' (inside $\F$, right of
$\U$, left of $\M$) and the ``$\phiS$-context'' (inside $\G$, right of
$\R$, left of $\W$), producing a formula of mixed class in each case.
The new translation in Section~\ref{sec:new} addresses this by
introducing two auxiliary functions that produce $\LTLG$ and $\LTLS$
formulas, respectively, and using them in the appropriate positions.

\section{The New Reduction from \texorpdfstring{\LTLf}{LTLf} to Syntactic Obligation}
\label{sec:new}

We define three mutually recursive functions $\tG$, $\tS$, and $\tO$,
supported by an auxiliary function $\tB$, with the following invariants
established by structural induction:
\begin{itemize}
\item For any \LTLf formula $f$, $\tG(f)\in\LTLG$ and $\tG(f)\eqalv\tv(f)$;
\item For any \LTLf formula $f$, $\tS(f)\in\LTLS$ and $\tS(f)\eqalv\tv(f)$;
\item For any \LTLf formula $f$, $\tO(f)\in\LTLO$ and $\tO(f)\eqalv\tv(f)$.
\end{itemize}

The auxiliary function $\tB$ is simply a restriction of $\tv$ to inputs
that are in $\LTLfB$.  So it also naturally follows that
\begin{itemize}
  \item If $f\in\LTLfB$ then $\tB(f)\in\LTLB$ and $\tB(f)\eqalv\tv(f)$
\end{itemize}

\subsection{Revised top-level wrapper}\label{sec:revisited-wrapper}

The wrapper $\alv \U \G\lnot\alv$ from \eqref{eq:wrapper-dv} is not in
$\LTLO$.  We replace it with an $\LTLO$ formula equivalent to it on alive-encoded words:
\[
  \F\lnot\alv \;\land\; (\alv \W \G\lnot\alv),
\]

The new wrapper is therefore:
\begin{equation}\label{eq:wrapper-new}
  \mathrm{from\_ltlf}_O(f) \;=\;
  \underbrace{\tO(f)}_{\in\LTLO} \;\land\; \underbrace{\alv}_{\in\LTLB} \;\land\; \underbrace{\F\lnot\alv}_{\in\LTLG} \;\land\; \underbrace{(\alv \W \G\lnot\alv)}_{\in\LTLS}.
\end{equation}
All four conjuncts belong to $\LTLO$ or its subclasses, so the full formula belongs to $\LTLO$.

\subsection{The function \texorpdfstring{$\tB$}{tB}}

The function $\tB$ translates formulas of $\LTLfB$ (the bottom class
extended with $\StrongX$) into formulas of $\LTLB$.  This restriction
of the translation uses the same rules as the original $\tv(\cdot)$ function.

\begin{alignat*}{2}
  \tB(\bot)       &= \bot, \quad \tB(\top) = \top, \quad \tB(v) = v \\
  \tB(\lnot f)    &= \lnot\tB(f)
                   &&\quad[\lnot\varphi_B \in \LTLB] \\
  \tB(f \odot g)  &= \tB(f) \odot \tB(g)\quad\text{ for any}~\odot \in \{\land,\lor,\to,\liff,\lxor\}
                   &&\quad[\varphi_B \odot \varphi_B \in \LTLB] \\
  \tB(\X f)       &= \X(\lnot\alv \lor \tB(f))
                   &&\quad[\X\varphi_B \in \LTLB] \\
  \tB(\StrongX f) &= \X(\alv \land \tB(f))
                   &&\quad[\X\varphi_B \in \LTLB]
\end{alignat*}

In the implementation, $\tB$ is used by $\tG$ and $\tS$ to avoid expanding
$\lxor$ and $\liff$ when their operands belong to $\LTLfB$.

\subsection{The function \texorpdfstring{$\tG$}{tG}}

The function $\tG$ produces a $\LTLG$ formula equivalent to $\tv(f)$ at
alive positions.  It converts every weak temporal operator into an equivalent
$\U$-based form (since $\U$ stays in $\LTLG$) and recurses with $\tS$
wherever a $\LTLS$ subformula is needed (e.g.\ as the negated argument or
the antecedent of an implication).

\begin{alignat*}{2}
  \tG(\bot)       &= \bot, \quad \tG(\top) = \top, \quad \tG(v) = v \\
  \tG(\lnot f)    &= \lnot\tS(f)
                   &&\quad[\lnot\varphi_S \in \LTLG] \\
  \tG(f \btriangle g) &= \tG(f) \btriangle \tG(g)\quad\text{ for any}~\triangle \in \{\land,\lor\}
                   &&\quad[\varphi_G \btriangle \varphi_G \in \LTLG] \\
  \tG(f \to g)    &= \tS(f) \to \tG(g)
                   &&\quad[\varphi_S \to \varphi_G \in \LTLG] \\
  \tG(f \lxor g)  &=
      \begin{cases}
        \tB(f \lxor g) & \text{if } f\lxor g\in\LTLfB\\[2mm]
        \bigl(\tG(f) \land \lnot\tS(g)\bigr)
        \lor \bigl(\lnot\tS(f) \land \tG(g)\bigr) & \text{otherwise}
      \end{cases}
                   &&\quad[\LTLG\text{ not closed under }\lxor] \\
  \tG(f \liff g)  &=
      \begin{cases}
        \tB(f \liff g) & \text{if } f\liff g\in\LTLfB\\[2mm]
        \bigl(\tG(f) \land \tG(g)\bigr)
        \lor \bigl(\lnot\tS(f) \land \lnot\tS(g)\bigr) & \text{otherwise}
      \end{cases}
                   &&\quad[\LTLG\text{ not closed under }\liff] \\
  \tG(\X f)       &= \X(\lnot\alv \lor \tG(f))
                   &&\quad[\X\varphi_G \in \LTLG] \\
  \tG(\StrongX f) &= \X(\alv \land \tG(f))
                   &&\quad[\X\varphi_G \in \LTLG] \\
  \tG(\F f)       &= \F(\alv \land \tG(f))
                   &&\quad[\F\varphi_G \in \LTLG] \\
  \tG(\G f)       &= \tG(f) \U \lnot\alv
                   &&\quad[\varphi_G \U \varphi_G \in \LTLG\text{; see below}] \\
  \tG(f \U g)     &= \tG(f) \U (\alv \land \tG(g))
                   &&\quad[\varphi_G \U \varphi_G \in \LTLG] \\
  \tG(f \M g)     &= (\alv \land \tG(f)) \M \tG(g)
                   &&\quad[\varphi_G \M \varphi_G \in \LTLG] \\
  \tG(f \R g)     &= \tG(g) \U (\lnot\alv \lor (\tG(f) \land \tG(g)))
                   &&\quad[\LTLG\text{; derived below}] \\
  \tG(f \W g)     &= \tG(f) \U (\lnot\alv \lor \tG(g))
                   &&\quad[\LTLG\text{; derived below}]
\end{alignat*}

\paragraph{Derivations.}
\begin{itemize}
  \item Under the alive prefix, $\tG(f) \U \lnot\alv \eqalv \G f$:
        the $\U$ holds iff $\tG(f)$ holds at every alive position (the
        ``exit'' branch of $\U$ is triggered at the first $\lnot\alv$ position).
        Hence $\tG(\G f) = \tG(f) \U \lnot\alv$.
  \item $f \W g \equiv (f \U g) \lor \G f$, so its encoding is
        \begin{align*}
          &(\tG(f) \U (\alv \land \tG(g))) \lor (\tG(f) \U \lnot\alv) \\
          &= \tG(f) \U ((\alv \land \tG(g)) \lor \lnot\alv)
           = \tG(f) \U (\lnot\alv \lor \tG(g)).
        \end{align*}
  \item $f \R g \equiv g \W (f \land g)$, giving
        $\tG(g) \U (\lnot\alv \lor (\tG(f) \land \tG(g)))$.
\end{itemize}

\subsection{The function \texorpdfstring{$\tS$}{tS}}

The function $\tS$ is the dual of $\tG$: it produces a $\LTLS$ formula
equivalent to $\tv(f)$ at alive positions.  It converts every strong temporal
operator into an $\R$-based form and recurses with $\tG$ wherever a $\LTLG$
subformula is needed.

\begin{alignat*}{2}
  \tS(\bot)       &= \bot, \quad \tS(\top) = \top, \quad \tS(v) = v \\
  \tS(\lnot f)    &= \lnot\tG(f)
                   &&\quad[\lnot\varphi_G \in \LTLS] \\
  \tS(f \btriangle g) &= \tS(f) \btriangle \tS(g)\quad\text{ for any}~\triangle \in \{\land,\lor\}
                   &&\quad[\varphi_S \btriangle \varphi_S \in \LTLS] \\
  \tS(f \to g)    &= \tG(f) \to \tS(g)
                   &&\quad[\varphi_G \to \varphi_S \in \LTLS] \\
  \tS(f \lxor g)  &=
      \begin{cases}
        \tB(f \lxor g) & \text{if } f\lxor g\in\LTLfB\\[2mm]
        \bigl(\tS(f) \land \lnot\tG(g)\bigr)
        \lor \bigl(\lnot\tG(f) \land \tS(g)\bigr) & \text{otherwise}
      \end{cases}
                   &&\quad[\LTLS\text{ not closed under }\lxor] \\
  \tS(f \liff g)  &=
      \begin{cases}
        \tB(f \liff g) & \text{if } f\liff g\in\LTLfB\\[2mm]
        \bigl(\tS(f) \land \tS(g)\bigr)
        \lor \bigl(\lnot\tG(f) \land \lnot\tG(g)\bigr) & \text{otherwise}
      \end{cases}
                   &&\quad[\LTLS\text{ not closed under }\liff] \\
  \tS(\X f)       &= \X(\lnot\alv \lor \tS(f))
                   &&\quad[\X\varphi_S \in \LTLS] \\
  \tS(\StrongX f) &= \X(\alv \land \tS(f))
                   &&\quad[\X\varphi_S \in \LTLS] \\
  \tS(\G f)       &= \G(\lnot\alv \lor \tS(f))
                   &&\quad[\G\varphi_S \in \LTLS] \\
  \tS(\F f)       &= \tS(f) \R \alv
                   &&\quad[\varphi_S \R \varphi_S \in \LTLS\text{; key case}] \\
  \tS(f \R g)     &= \tS(f) \R (\lnot\alv \lor \tS(g))
                   &&\quad[\varphi_S \R \varphi_S \in \LTLS] \\
  \tS(f \W g)     &= (\lnot\alv \lor \tS(f)) \W \tS(g)
                   &&\quad[\varphi_S \W \varphi_S \in \LTLS] \\
  \tS(f \U g)     &= \tS(g) \R (\alv \land (\tS(f) \lor \tS(g)))
                   &&\quad[\LTLS\text{; derived below}] \\
  \tS(f \M g)     &= \tS(f) \R (\alv \land \tS(g))
                   &&\quad[\LTLS\text{; derived below}]
\end{alignat*}

\paragraph{Derivations.}
\begin{itemize}
  \item Under the alive prefix, $\F(\alv \land \tS(f)) \eqalv \tS(f) \R \alv$:
        the $\R$ holds iff $\tS(f)$ holds at some alive position (the
        ``forever'' branch of $\R$ is ruled out by $\alv$ eventually becoming
        false).  Hence $\tS(\F f) = \tS(f) \R \alv$.
  \item $f \U g \equiv \lnot((\lnot f) \R (\lnot g))$, so by duality with
        $\tG$: expanding with the $\tG$ rule for $\R$ and simplifying gives
        $\tS(f \U g) = \tS(g) \R (\alv \land (\tS(f) \lor \tS(g)))$.
  \item $f \M g \equiv \lnot((\lnot f) \W (\lnot g))$, similarly yielding
        $\tS(f) \R (\alv \land \tS(g))$.
\end{itemize}

\subsection{The function \texorpdfstring{$\tO$}{tO}}

The function $\tO$ is the main translation.  Boolean operators are treated
homomorphically (since $\LTLO$ is closed under all Boolean connectives).
Temporal operators use $\tG$ or $\tS$ on the operand that requires a
constrained syntactic class, and $\tO$ recursively on the unconstrained
operand.

\begin{alignat*}{2}
  \tO(\bot)       &= \bot, \quad \tO(\top) = \top, \quad \tO(v) = v \\
  \tO(\lnot f)    &= \lnot\tO(f)
                   &&\quad[\lnot\varphi_O \in \LTLO] \\
  \tO(f \odot g)  &= \tO(f) \odot \tO(g)\quad\text{ for any}~\odot \in \{\land,\lor,\to,\liff,\lxor\}
                   &&\quad[\varphi_O \odot \varphi_O \in \LTLO] \\
  \tO(\X f)       &= \X(\lnot\alv \lor \tO(f))
                   &&\quad[\X\varphi_O \in \LTLO] \\
  \tO(\StrongX f) &= \X(\alv \land \tO(f))
                   &&\quad[\X\varphi_O \in \LTLO] \\
  \tO(\F f)       &= \F(\alv \land \tG(f))
                   &&\quad[\F\varphi_G \in \LTLG \subseteq \LTLO] \\
  \tO(\G f)       &= \G(\lnot\alv \lor \tS(f))
                   &&\quad[\G\varphi_S \in \LTLS \subseteq \LTLO] \\
  \tO(f \U g)     &= \tO(f) \U (\alv \land \tG(g))
                   &&\quad[\varphi_O \U \varphi_G \in \LTLO] \\
  \tO(f \R g)     &= \tO(f) \R (\lnot\alv \lor \tS(g))
                   &&\quad[\varphi_O \R \varphi_S \in \LTLO] \\
  \tO(f \W g)     &= (\lnot\alv \lor \tS(f)) \W \tO(g)
                   &&\quad[\varphi_S \W \varphi_O \in \LTLO] \\
  \tO(f \M g)     &= (\alv \land \tG(f)) \M \tO(g)
                   &&\quad[\varphi_G \M \varphi_O \in \LTLO]
\end{alignat*}

\begin{example}
  Revisiting Example~\ref{ex:gfa}:
  \begin{align*}
    \tO(\G(\F a)) &= \G\bigl(\lnot\alv \lor \tS(\F a)\bigr)
                   = \G\bigl(\lnot\alv \lor (a \R \alv)\bigr).
  \end{align*}
  Here, the whole formula is in $\LTLS \subseteq \LTLO$.
  \begin{align*}
    \tO(\F(\G a)) &= \F\bigl(\alv \land \tG(\G a)\bigr)
                   = \F\bigl(\alv \land (a \U \lnot\alv)\bigr).
  \end{align*}
  Here, the whole formula is in $\LTLG \subseteq \LTLO$.
\end{example}

\subsection{Correctness}

\paragraph{Syntactic-class membership.}
By structural induction: every rule for $\tB$ produces a formula in $\LTLB$,
every rule for $\tG$ produces a formula in $\LTLG$,
every rule for $\tS$ produces a formula in $\LTLS$, and every rule for $\tO$
produces a formula in $\LTLO$, as annotated above.  The only non-obvious
cases are $\lxor$ and $\liff$ in $\tG$ and $\tS$: because $\LTLG$ (respectively
$\LTLS$) is closed under $\lxor$ and $\liff$ only when both operands are in
$\LTLfB$, we delegate to $\tB$ for $\LTLfB$ formulas.  Otherwise, we expand
those connectives into $\land/\lor/\lnot$ combinations of $\tG$ and $\tS$
results, which \emph{are} closed under $\land$ and $\lor$.

\paragraph{Proof sketch: semantic equivalence.}
By simultaneous structural induction on $f$, with all formulas evaluated on an alive-encoded word at an alive position, covering $\tG$, $\tS$, and $\tO$
(and their mutual recursion), we show:
\begin{itemize}
  \item \emph{$\tG(f) \eqalv \tv(f)$}:
        the key case is $\G f$.
        At alive position $i$, $\tG(f) \U \lnot\alv$ holds iff $\tG(f)$
        (which agrees with $\tv(f)$ by induction) holds at all alive positions
        $j \geq i$, which is exactly the meaning of $\G(\lnot\alv \lor \tv(f))$
        at $i$.
  \item \emph{$\tS(f) \eqalv \tv(f)$}:
        the key case is $\F f$.
        At alive position $i$, $\tS(f) \R \alv$ holds iff $\tS(f)$
        holds at some alive position $j \geq i$ (the ``forever'' branch of $\R$
        is ruled out by the live prefix being finite), matching
        $\F(\alv \land \tv(f))$.
  \item \emph{$\tO(f) \eqalv \tv(f)$}:
        each mixed temporal rule uses $\tG$ (resp.\ $\tS$) on the constrained
        operand; by the above, those agree with $\tv$ at alive positions, so
        $\tO(f)$ and $\tv(f)$ agree at every alive position.
\end{itemize}
The invariant for $\tB$ follows because $\tB$ is simply the restriction of
$\tv$ to inputs in $\LTLfB$: since $\tB$ uses the same homomorphic rules
as $\tv$ for the common operators, $\tB(f)\eqalv\tv(f)$ holds for
$f\in\LTLfB$ by the same structural induction.

Together with the corrected wrapper \eqref{eq:wrapper-new}, the new
translation therefore has the same satisfiability behavior as the original
on alive-encoded words.

\subsection{Why Replace $\tv$ by Four Functions?}

As mentioned at the top of section~\ref{sec:new}, $\tv(f)$, $\tS(f)$,
$\tG(f)$, and $\tO(f)$ all produce formulas that are equivalent under
the alive prefix.   For instance:
\begin{align*}
  \tv(\F\G a) &= \F\bigl(\alv \land \G(\lnot\alv \lor a)\bigr) & [\not\in& \LTLO]\\
  \tO(\F\G a) = \tG(\F\G a) &= \F\bigl(\alv \land (a \U \lnot\alv)\bigr) & [\in&\LTLG]\\
  \tS(\F\G a) &= \G(\lnot\alv \lor a)\R \alv & [\in&\LTLS]\\
\end{align*}

These LTL formulas are not equivalent in general, but are equivalent
at the $\alv$ position of alive-encoded words.

The wrapper \eqref{eq:wrapper-new} adds a combination of terms that
are in $\LTLS$ and in $\LTLG$ in every case, so choosing $\tO$, $\tS$,
or $\tG$ does not change the class of the resulting formula.

The reason we prefer to use $\tO$ is that it will change fewer
temporal operators when possible (e.g., if the \LTLf formula already syntactically belongs to $\LTLO$, it will only add the $\alv$
proposition, as in the original $\tv$ function).  Furthermore,
compared to $\tS$ and $\tG$, it can also deal with $\lxor$ and $\liff$
naturally.

Function $\tB$ was added as an optimization to recover subcases where $\lxor$ and $\liff$ can be supported without blowup.

\section{Availability in Spot}

The new $\mathrm{from\_ltlf}_O$ function has been implemented in Spot,
and is intended to be available in the upcoming Spot 2.16 release:

\begin{verbatim}
% ltlfilt --from-ltlf -f 'GFa'
alive & G(!alive | (a R alive)) & F!alive & (alive W G!alive)
% ltlfilt --from-ltlf -f 'GFa' --syntactic-obligation -c
1
\end{verbatim}

In Spot 2.16, the old reduction, which does not produce syntactic
obligations, can still be obtained by setting the
\verb|SPOT_FROM_LTLF| environment variable to \texttt{0}.

\begin{verbatim}
% export SPOT_FROM_LTLF=0
% ltlfilt --from-ltlf -f 'GFa'
alive & G(!alive | F(a & alive)) & (alive U G!alive)
% ltlfilt --from-ltlf -f 'GFa' --syntactic-obligation -c
0
\end{verbatim}

On this last example, the third command shows that the result of the
second one is not a syntactic obligation.

\bibliographystyle{plainnat}
\bibliography{spot}

\end{document}